# Integration of p-type $Cr_2O_3$ on Ultra-Wide Bandgap AlGaN PolFETs with 2.5 kV Breakdown Voltage

Jonathan Pratt[1,a)], Yizheng Liu[3], Ijaas Mohamed[1], Seungheon Shin[1], Akilesh Srikanth[1], Md Saklain Morshed[1], Brianna A. Klein[4], Andrew Armstrong[4], Andrew A. Allerman[4], Sriram Krishnamoorthy[3], and Siddharth Rajan[1,2]

[1]*Department of Electrical and Computer Engineering, Ohio State University, Columbus, OH, USA*

[2]*Department of Materials Science and Engineering, Ohio State University, Columbus, OH, USA*

[3]*Materials Department, University of California, Santa Barbara, CA, USA*

[4]*Sandia National Laboratories, Albuquerque, NM, USA*

***Abstract*—We demonstrate UWBG AlGaN polarization-graded field-effect transistors (PolFETs) incorporating room-temperature sputtered p-type oxide $Cr_2O_3$ as a P-N heterojunction gate, enabling superior field management for lateral device breakdown compared to conventional Schottky gates. Devices using reverse-graded n-AlGaN contact layers achieved a record-low $R_c$ of 0.56 Ω·mm. The $Cr_2O_3$ gate exhibits a turn-on voltage $V_{ON}$ > 3.5 V and a positive threshold voltage shift of +1.63 V relative to Schottky gates, displaying enhanced channel depletion from the P-N junction. Fabricated devices show high $I_{max}$ (590 mA/mm) and $I_{ON}/I_{OFF}$ of $3\times10^7$. Devices exhibited state-of-the-art $V_{BR}$ > 2.5 kV with 11.2 mΩ·cm² specific on-resistance ($L_{GD}$ = 9.55 µm, average breakdown field > 2.5 MV/cm), while shorter gate-drain devices showed high breakdown fields up to 5.3 MV/cm with 0.28 mΩ·cm² specific on-resistance. These results showcase sputtered p-$Cr_2O_3$ as a viable, low thermal budget P-N junction gate technology for high-Al-composition AlGaN transistors in RF and power electronics.**

---

a) Author to whom correspondence should be addressed

E-mail: *pratt.335@osu.edu*

Ultra-wide bandgap (UWBG) semiconductor materials are the subject of an increasing number of investigations for next-generation electronic devices. The aptly named ultra-wide bandgaps of UWBG materials like AlN, high Al-composition AlGaN, $Ga_2O_3$, diamond, and c-BN promise high-voltage blocking and critical electric fields surpassing wide bandgap materials like GaN and SiC, which represent the current commercial standards for lateral and vertical device technology. UWBG AlGaN (also known as Al-rich, high Al-content, and high Al-composition AlGaN) is a potential successor to GaN for both radio-frequency (RF) and lateral power transistors due to its comparable high saturation velocity ($\sim 2\times10^7$ cm/s) and its theoretical breakdown field (> 10 MV/cm). The lateral power figure of merit (LFOM) for AlGaN predicts superior performance during high-temperature operation due to strong phonon-driven mobility degradation in GaN[1].

Proper gate design and field management are necessary in lateral devices to achieve the theoretical fields across the gate-drain region while avoiding device breakdown. Gate dielectrics are a common solution to prevent breakdown at lower voltages due to barrier lowering and electron tunneling, which inhibit Schottky-gated devices. However, dielectrics likewise limit breakdown fields due to trapping effects and low intrinsic critical fields compared to typical UWBG materials. P-N junctions alternatively support large vertical fields, nearing theoretical limits for UWBG materials. Reverse-graded polarization-doped AlGaN and p-AlGaN have been used for the p-type junction side in vertical diodes[2,3]. For lateral devices, using P-N junctions in place of typical Schottky gates has the added benefit of depleting channel charge for enhancement-mode operation[4–6]. The p-type material has been extended into the gate-drain access regions in lateral GaN power devices to form superjunction structures for field management[7].

Incorporating p-type polarization-doped AlGaN or p-GaN in UWBG AlGaN access and gate regions is possible through two different design approaches to avoid high contact resistances. The first is through continuous epitaxial growth following growth of the channel, commonly performed with metal-organic chemical vapor deposition (MOCVD) or molecular-beam epitaxy (MBE). Regrown degenerately-doped n-GaN contacts are later formed to contact the lower Al-composition channel in etched ohmic regions[8,9]. The second approach is through continuously growing epitaxial reverse-graded n-AlGaN contacts atop the channel[10–14]. The contact layers are later etched to form the device access region. Patterned p-type material is then deposited in the etched region to form the P-N junction. Devices with reverse-graded contact layers have shown the lowest contact resistance to UWBG AlGaN to date[13].

The epitaxial integration of native p-type III-nitride material with reverse-graded contacts requires condition-sensitive, patterned regrowth at elevated temperatures, constraining fabrication processes. An alternative is to use sputtered p-type materials. Sputtered p-type oxides like $NiO_x$ have been incorporated into $Ga_2O_3$-based heterojunctions[15–20]. However, $NiO_x$ has severe electrical degradation in $H_2O$-ambient[21], limiting its commercial applications. Reports have demonstrated that p-type oxide $Cr_2O_3$ is a suitable alternative to $NiO_x$[22,23], where $Ga_2O_3$ films incorporating p-$Cr_2O_3$ achieved breakdown voltages exceeding 3.45 kV[24]. $Cr_2O_3$ films patterned with photoresist have been sputtered at room temperature, removing additional processing steps required for hard masks like $SiO_2$ or $SiN_x$ and reducing concerns of unintentional thermal annealing.

In this report, we demonstrate room-temperature sputtered p-type $Cr_2O_3$ incorporated on UWBG AlGaN polarization-graded field-effect transistors (PolFETs) to form P-N junction gates. Devices with p-$Cr_2O_3$ gates show lateral breakdown voltages up to 2562 V with $R_{ON,sp}$ = 11.2 mΩ·cm$^2$ ($L_{GD}$ = 9.55 µm). Short dimension devices show average breakdown electric fields of 5.3 MV/cm with maximum on-current, $I_{max}$, exceeding 590 mA/mm and $R_{ON,sp}$ = 0.28 mΩ·cm$^2$ ($L_{GD}$ = 0.55 µm). Gates formed with p-$Cr_2O_3$ show turn-on voltage ($V_{ON}$) above +3.5 V—near the DFT-predicted bandgap of $Cr_2O_3$[21] (3.45 eV)—and a positive threshold voltage shift of $\Delta V_{TH}$ = +1.63 V. PolFET devices with reverse-graded contact layers show record-low contact resistance of 0.56 Ω·mm. Results reported herein highlight the potential of p-type oxide gates for next-generation UWBG AlGaN heterostructure devices for power and RF electronics.

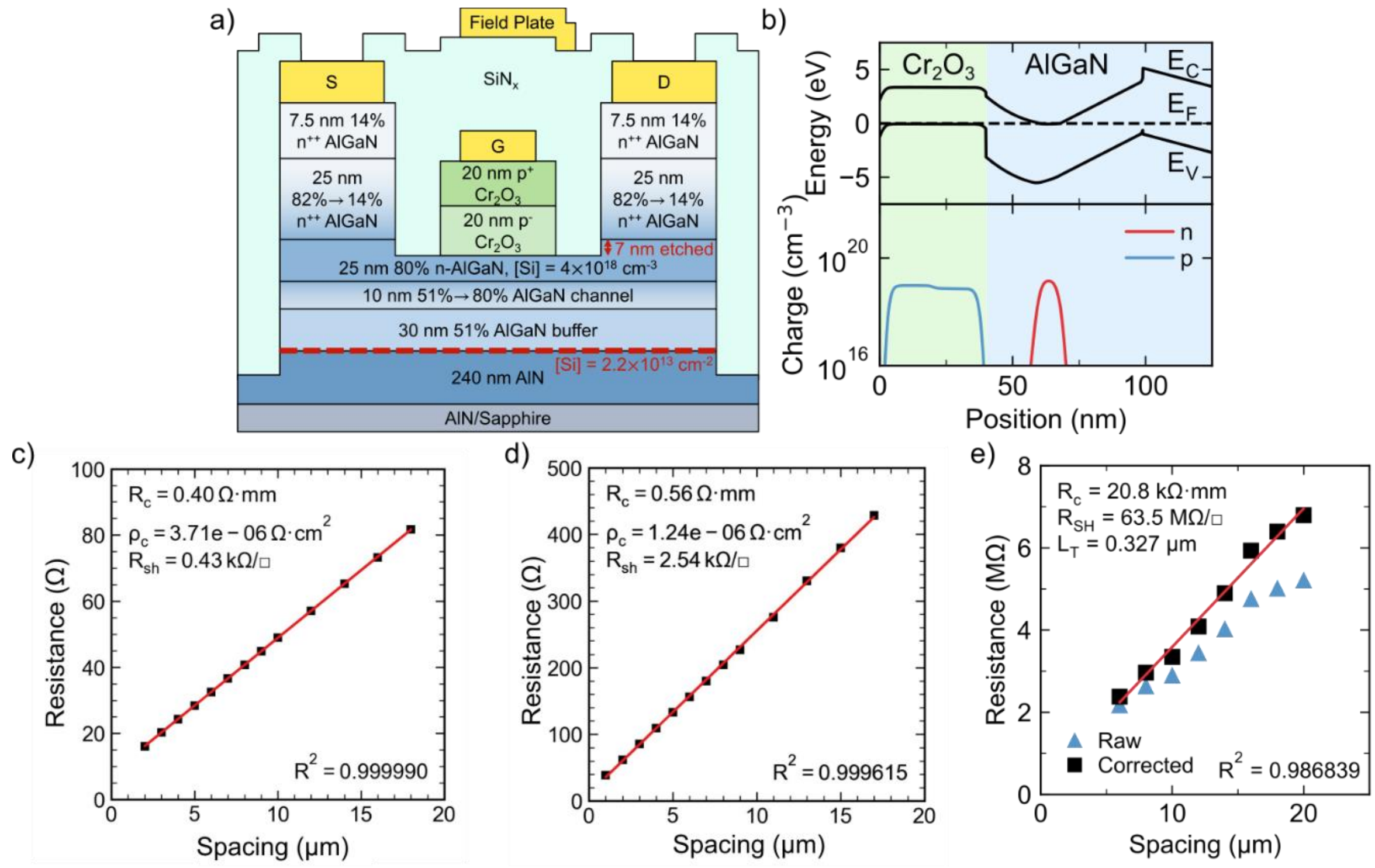

Figure 1. (a) Schematic of fabricated PolFET device structure, (b) simulated energy band diagram with $Cr_2O_3$ under the gate metal, (c) TLM measurements between patterns without etched n-AlGaN contact layers, (d) TLM measurements between patterns with etched n-AlGaN contact layers, (e) circular TLM measurement of $Cr_2O_3$. (Blue triangles: uncorrected raw circular TLM data, Black squares: geometry corrected circular TLM data).

AlGaN polarization-graded transistors were epitaxially grown with a Taiyo Nippon Sanso SR4000-HT MOCVD reactor using an AlN on sapphire template as a substrate. The epitaxial heterostructure (Fig. 1(a)) consists of a 240 nm AlN buffer, a 1 nm Si $\delta$-doped ($2.2\times10^{13}$ cm$^{-2}$) $Al_{0.51}Ga_{0.49}N$ back barrier, a 30 nm $Al_{0.51}Ga_{0.49}N$ spacer, a 10 nm $Al_{0.51}Ga_{0.49}N$ to $Al_{0.8}Ga_{0.2}N$ graded channel, a 25 nm n-$Al_{0.8}Ga_{0.2}N$ barrier and a reverse-graded n++ $Al_{0.82}Ga_{0.18}N$ to $Al_{0.14}Ga_{0.86}N$ contact layer. The heterostructure energy band diagram is shown in Fig. 1(b). Access regions were defined with low-damage ICP-RIE etching of the contact layer, resulting in a post-etch barrier thickness of 18 nm. ICP-RIE was used for mesa isolation. Hall measurements were taken using Van der Pauw pattern geometry using the reverse graded contacts to access the recessed AlGaN channel. A mobility of 223 cm$^2$/Vs, carrier concentration of $1.36\times10^{13}$ cm$^{-2}$, and sheet resistance of 2.05 kΩ/□ were measured on the bare AlGaN PolFET sample (without $Cr_2O_3$). Contact resistances and sheet resistances were measured using the transfer length method (TLM) on mesa isolated regions with and without the low-damage contact layer etching between TLM contacts. The TLM results in the region without contact layer etching (Fig. 1(c)) showed contact resistance $R_c = 0.40$ Ω·mm ($\rho_c = 3.71\times10^{-6}$ Ω·cm$^2$) and sheet resistance $R_{sh} = 0.43$ kΩ/□. The TLM results in the region with contact layer etching (Fig. 1(d)) showed contact resistance $R_c = 0.56$ Ω·mm ($\rho_c = 1.24\times10^{-6}$ Ω·cm$^2$), the lowest value reported to date, and sheet resistance $R_{sh} = 2.54$ kΩ/□.

Two sets of transistors were fabricated. In the first set (called "Schottky gate"), Ni/Au (30/100 nm) Schottky gates were deposited directly on the AlGaN barrier layer. In the second set (called "$Cr_2O_3$ gate"), gate patterns were defined using bilayer photoresist and direct-write optical lithography. A patterned bilayer $Cr_2O_3$ film consisting of 20 nm of $p^-$ $Cr_2O_3$ followed by 20 nm of $p^+$ $Cr_2O_3$ was deposited at room temperature by RF reactive magnetron sputtering[21], followed by Ni/Au (30/100 nm) using e-beam evaporation. The $Cr_2O_3$ and Ni/Au layers were then lifted off. Ti/Al/Ni/Au (20/120/30/100 nm) contacts were deposited in the ohmic regions to define the source and drain followed by Ni/Au (30/100 nm) bond pads. The devices were passivated with 150 nm of $SiN_x$ using plasma-enhanced chemical vapor deposition (PECVD). Gate-connected field plates were formed with Ni/Au/Ni (30/100/20 nm).

To measure the properties of the $Cr_2O_3$ layer, the same 20 nm $p^+$ $Cr_2O_3$/20 nm $p^-$ $Cr_2O_3$ deposition was done on a control sample (n-$Al_{0.3}Ga_{0.7}N$/AlN/sapphire). Ni/Au (30/100 nm) was deposited on the sample to form circular geometry TLM patterns. Four-terminal circular TLM was measured, and resistances were extracted from an I–V region below the P-N junction turn-on (–300 mV to +300 mV). The $Cr_2O_3$ circular TLM results (Fig. 1(e)) showed a sheet resistance $R_{sh}$ = 63.5 MΩ/□. Hall measurements were taken using Van der Pauw pattern geometry. A sheet resistance $R_{sh}$ = 8.8 MΩ/□ was extracted from the Hall measurements.

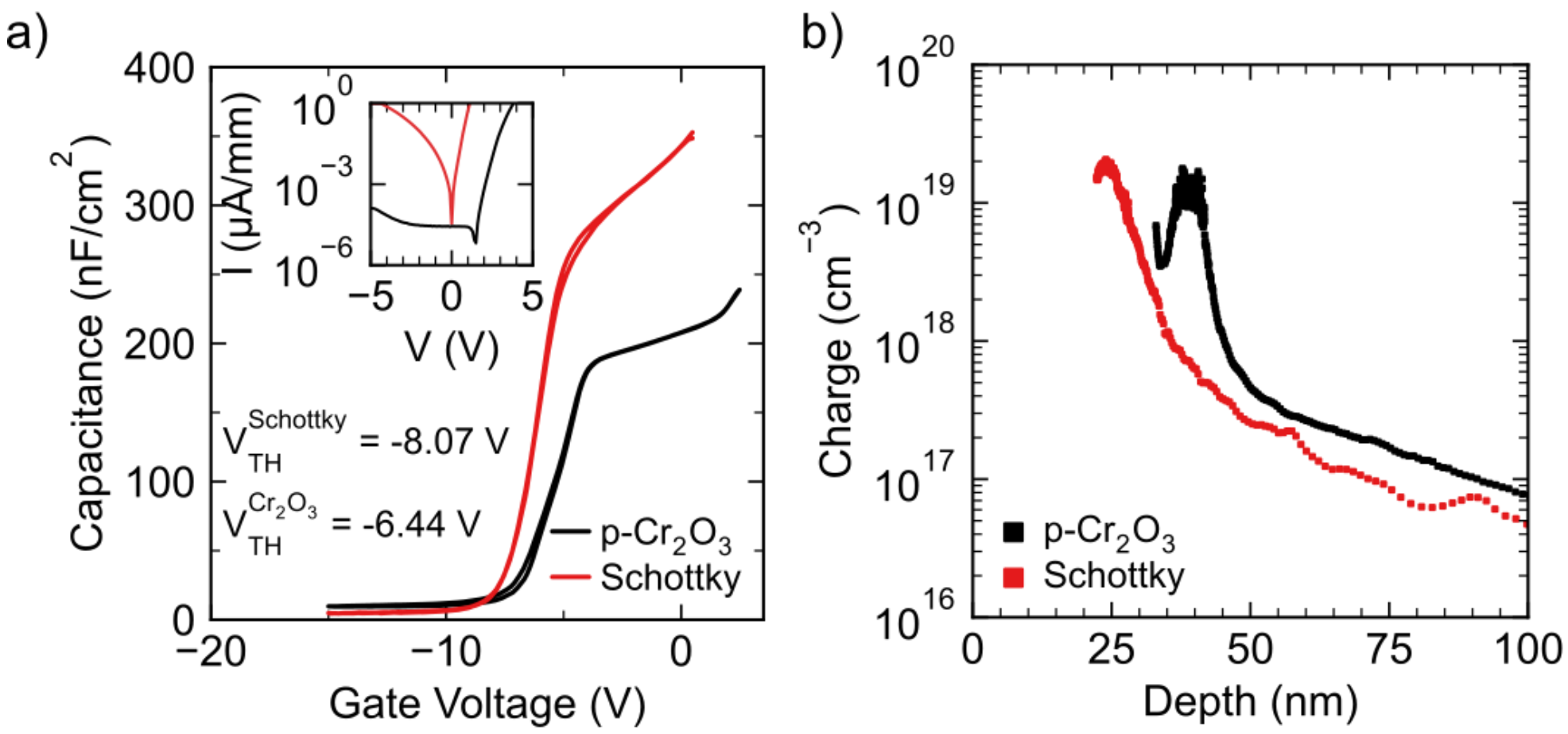

Figure 2. (a) Measured double-sweep C–V of devices at 1 MHz with $Cr_2O_3$ and Schottky gates (inset: 2-terminal I–V measurement of both gates), (b) extracted charge profiles from C–V.

Capacitance–voltage (*C–V*) measurements were taken for both Schottky and $Cr_2O_3$-gated devices at 1 MHz (Fig. 2(a)). The devices were measured from –15 V to below their respective turn-on voltages $V_{ON}^{Schottky}$ = 0.5 V and $V_{ON}^{Cr_2O_3}$ = 3.5 V. Threshold voltages were extracted at $V_{TH}^{Schottky}$ = –8.07 V and $V_{TH}^{Cr_2O_3}$ = –6.44 V, a positive threshold shift of $\Delta V_{TH}$ = +1.63 V. The positive threshold shift is evidence of a P-N heterojunction depleting additional charge in the barrier/channel compared to the Schottky gate. The difference in threshold voltage can be explained from the energy band diagram in Fig. 1(b). The total built-in voltage at equilibrium for the $Cr_2O_3$/AlGaN structure is ~ 3.5 eV, and the equilibrium band-bending in the n-$Al_{0.8}Ga_{0.2}N$ was estimated at 2.53 eV. In the Ni/Au Schottky, the built-in voltage is approximately 1.1 eV in UWBG AlGaN[25]. Thus, there is approximately 1.43 eV more band-bending in the $Cr_2O_3$/AlGaN device when compared with a Ni/AlGaN structure. This extra band-bending explains the shift in threshold voltage of ~ 1.63 V. Integrated zero-bias charge densities of $1.17\times10^{13}$ cm$^{-2}$ and $0.74\times10^{13}$ cm$^{-2}$ were respectively estimated from devices with Schottky and $Cr_2O_3$ gates. The energy band diagram and charge-depth profile (Fig. 1(b)) were simulated for the $Cr_2O_3$-gated device using a 1-D Schrodinger-Poisson solver[26]. We assumed an acceptor concentration of $N_A = 6\times10^{19}$ cm$^{-3}$ and acceptor ionization energy of 0.1 eV[27] for the $Cr_2O_3$. At zero bias, the simulated P-N depletion widths in the $Cr_2O_3$ and n-$Al_{0.8}Ga_{0.2}N$ were respectively estimated at 1.6 nm and 18 nm, indicating full depletion of the n-$Al_{0.8}Ga_{0.2}N$ barrier and a relatively low depletion width in the $Cr_2O_3$. The simulated charge density at zero-bias for $Cr_2O_3$ was $7.96 \times 10^{12}$ cm$^{-2}$, which is in good agreement with the measured C-V integrated charge density at 0 V ($7.3 \times 10^{12}$ cm$^{-2}$).

The Schottky-gate charge profile (Fig. 2(b)) has a visible peak with nominal ~$2\times10^{19}$ cm$^{-3}$ charge associated with the 10 nm graded channel. The $Cr_2O_3$-gated charge profile (Fig. 2(b)) displays a similar channel peak but contrasts with a lower nominal charge of ~$1.5\times10^{19}$ cm$^{-3}$, fully profiled and offset from the Schottky-gated channel peak by ~15 nm. The extracted charge profile in the device with a $Cr_2O_3$ gate is the result of dual depletion from the top p-type gate and the bottom n-type barrier/channel. The decrease in charge magnitude corresponds to the two-sided depletion, where the measured charge profile is $N^{-1} = N_D^{-1} + N_A^{-1}$. An estimated acceptor doping concentration of $N_A = 6\times10^{19}$ cm$^{-3}$ accounts for the decreased charge, used to simulate the band diagram in Fig. 1(b). The profile's depth offset compared to the Schottky gate is believed to be associated with the depletion width introduced in the $Cr_2O_3$ at the $Cr_2O_3$/n-AlGaN interface due to deep acceptors. High doping with deep acceptors introduces a region where deep acceptors are fully ionized in deep depletion but not ionized in shallow depletion[28]. Charge balance with the barrier and channel, comprising the n-side of the P-N junction, at zero bias induces p-side depletion. As bias is applied to the gate, ionization of the high nominal $Cr_2O_3$ doping charge balances depletion of charge in the n-AlGaN without a large change in the p-side depletion width, resulting in a near-fixed depletion width in the $Cr_2O_3$.

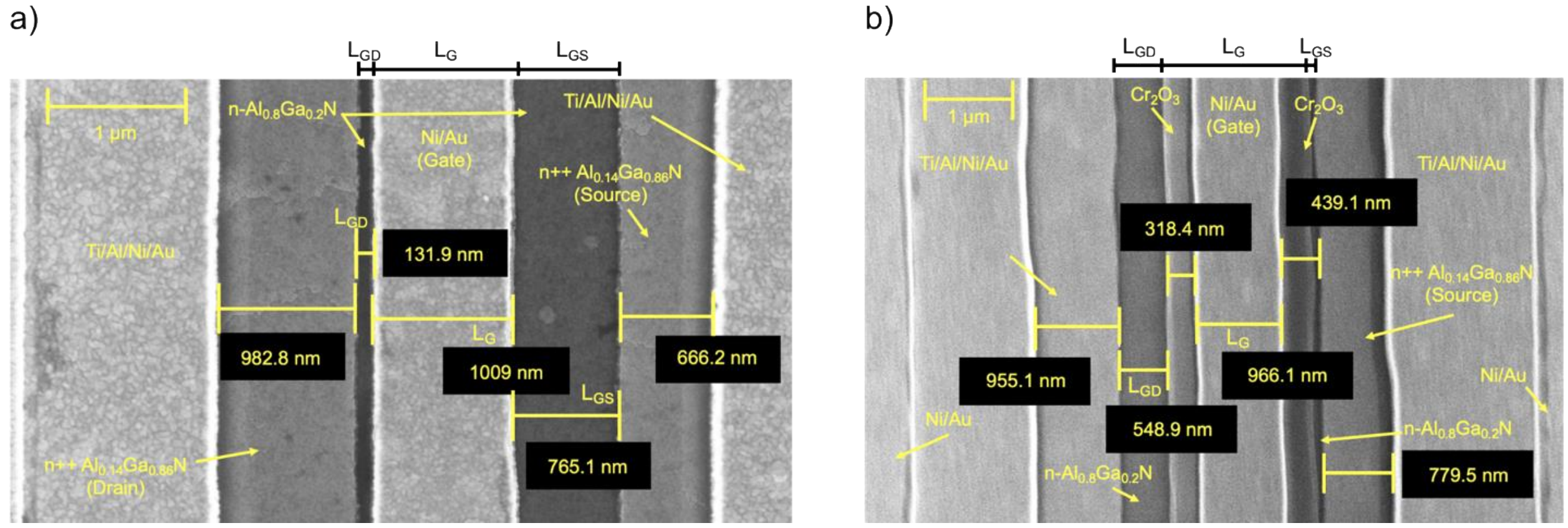

Figure 3. SEM imaging of fabricated PolFET devices with (a) Schottky gates and (b) $Cr_2O_3$ gates with visible extensions of the conformally sputtered $Cr_2O_3$ beyond the gate metal.

Table 1. Physical dimensions of Schottky gate and $Cr_2O_3$ gate devices measured with SEM.

| Device Gate | $L_{GD}$ (μm) | $L_G$ (μm) | $L_{GS}$ (μm) |
|---|---|---|---|
| Schottky | 0.132 | 1.01 | 0.765 |
| $Cr_2O_3$ | 0.867 | 0.966 | 0.439 |

Physical dimensions of selected Schottky-gated (Fig. 3(a)) and $Cr_2O_3$-gated (Fig. 3(b)) transistors measured with scanning electron microscopy (SEM) are tabulated in Table 1. Dimensions measured in SEM include the gate-drain spacings ($L_{GD}$), gate lengths ($L_G$), and gate-source spacings ($L_{GS}$). The SEM shows conformal deposition of the sputtered $Cr_2O_3$ on drain and source sides of the gate, where $Cr_2O_3$ was able to deposit within the photoresist bilayer's mask undercut. No extensions are visible in SEM on the selected Schottky-gated device. The $Cr_2O_3$ extensions are estimated at approximately 300 nm towards the drain and source. These extensions expand the gate metal's electrostatic control of the channel to their edges, shrinking $L_{GD}$ and $L_{GS}$. Device dimensions mentioned henceforth will refer to the spacings that include gate extension and misalignment effects.

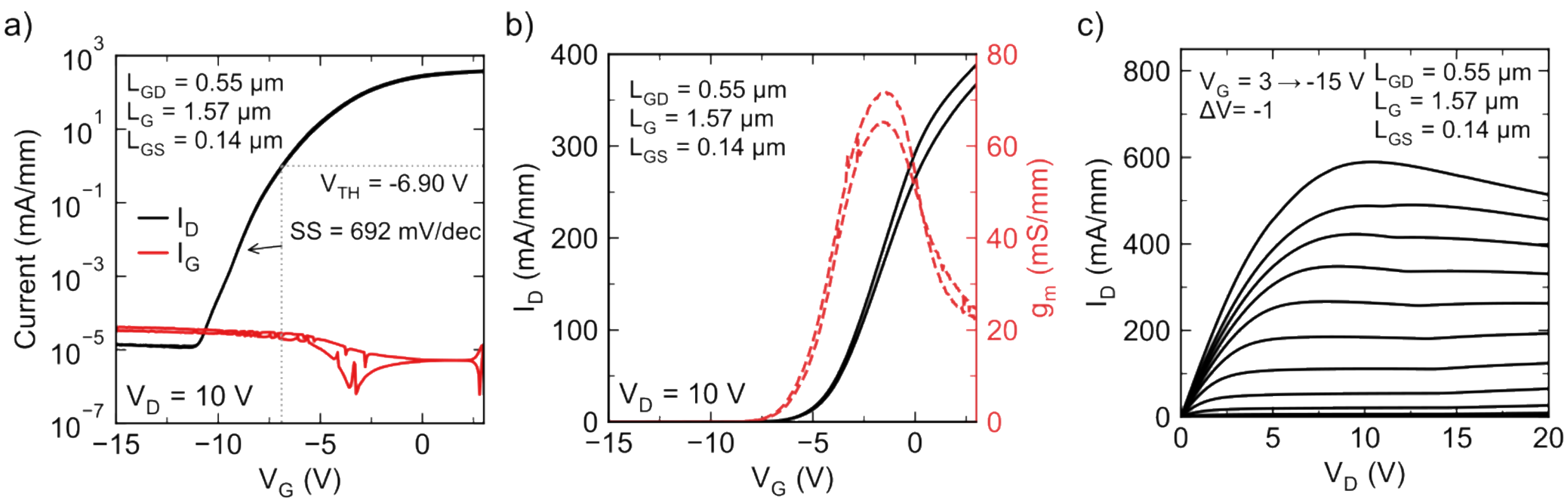

Figure 4. (a) Log-scale and (b) linear-scale transfer curves of fabricated device with $Cr_2O_3$ gate, (c) DC output characteristics of fabricated device with $Cr_2O_3$ gate.

Current-voltage ($I$–$V$) measurements of devices with a $Cr_2O_3$ gate were carried out under DC bias conditions. A pinch-off voltage, $V_p$, of –11 V was extracted from the $I$–$V$ transfer characteristics (Fig. 4(a)) using a drain voltage $V_D$ = 10 V, where devices showed a leakage current of $1.37\times10^{-5}$ mA/mm and an on-off current ratio $I_{ON}/I_{OFF}$ of $3\times10^{7}$. A threshold voltage ($V_{TH}$) of –6.90 V at $I_D$ = 1 mA/mm and a subthreshold slope ($SS$) of 692 mV/dec were extracted. A transconductance, $g_m$, exceeding 70 mS/mm was extracted (Fig. 4(b)). An anticlockwise hysteresis was observed in the $I$–$V$ measurements in both devices in the on-state, which we attribute to traps in the back-barrier. Output characteristics were measured up to gate voltage $V_G$ = 3 V (Fig. 4(c)). Devices with a $Cr_2O_3$ gate achieved a maximum on-state current, $I_{max}$, of 590 mA/mm for a source-to-drain spacing $L_{SD}$ = 2.26 µm. The specific on-resistance ($R_{ON,sp} = R_{ON} \times (2L_T + L_{SD})$) was extracted from the linear region of the output characteristics. Resistances $R_{ON,sp}$ = 0.28 mΩ·cm$^2$, 11.2 mΩ·cm$^2$, and 8.41 mΩ·cm$^2$ were respectively measured at gate-drain spacings $L_{GD}$ = 0.55 µm, 9.55 µm, and 14.55 µm at conditions: $V_G$ = 3 V, $V_D$ = 1 V.

Pulsed $I$–$V$ measurements were performed on a device with the dimensions $L_{GD}$ = 0.55 µm, $L_G$ = 1.57 µm, and $L_{GS}$ = 0.14 µm (Fig. 5). Measurements used quiescent voltages $V_{GSQ} = V_P - 3$ V and $V_{DSQ}$ = 40 V with a 6 µs pulse width, 0.1 % duty cycle. $V_{GSQ}$ = 0 V and $V_{DSQ}$ = 0 V were used for the baseline pulsed-IV condition. On-resistances were using the conditions mentioned above. $R_{ON}$ degradation between the two pulsed-IV conditions was 2.2%.

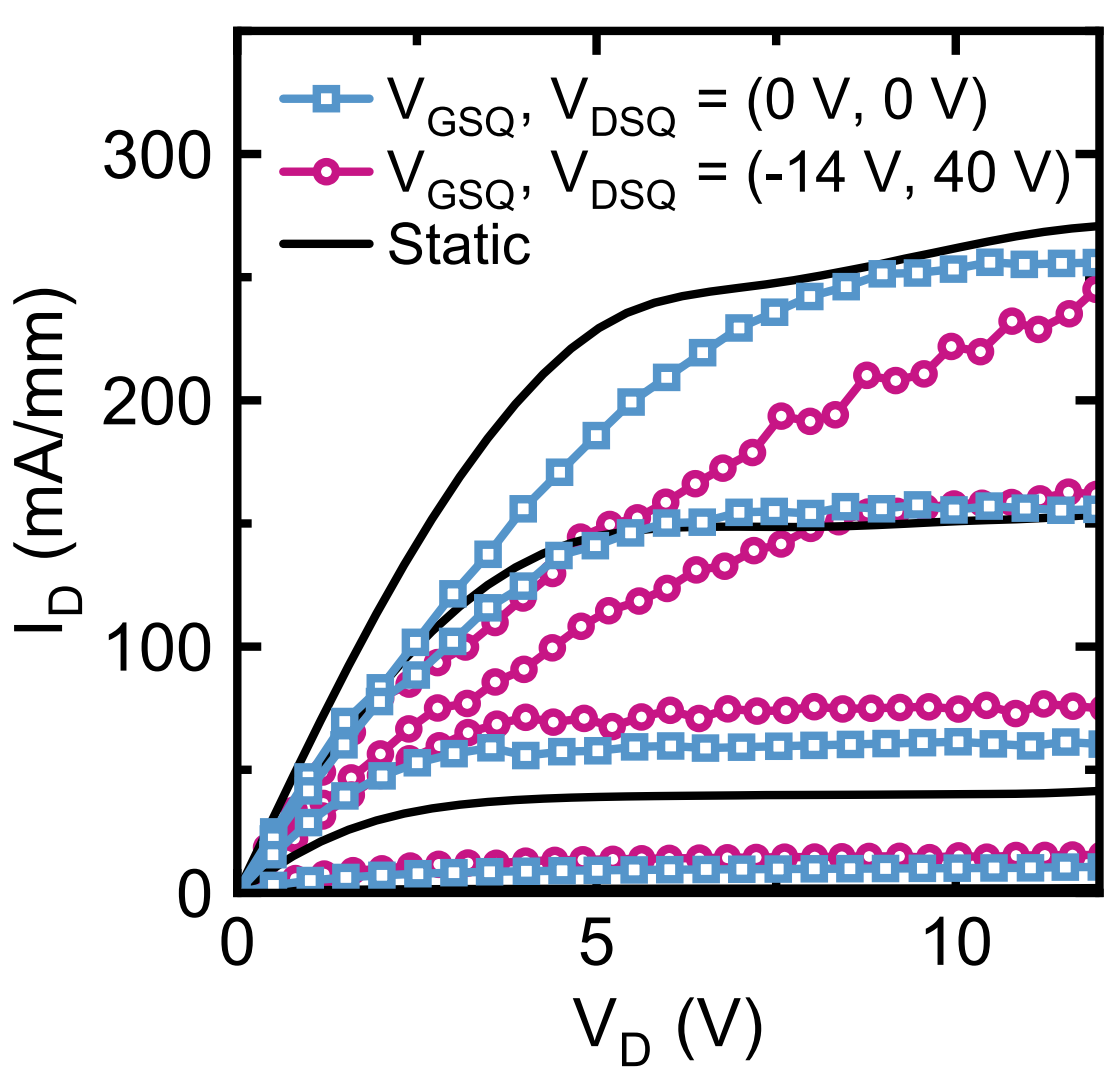

Figure 5. Pulsed–IV output characteristic measurements of device with $Cr_2O_3$ gate and $SiN_x$ passivation taken at $V_G$ = 3 to -6 V with step $\Delta V$ = -3 V.

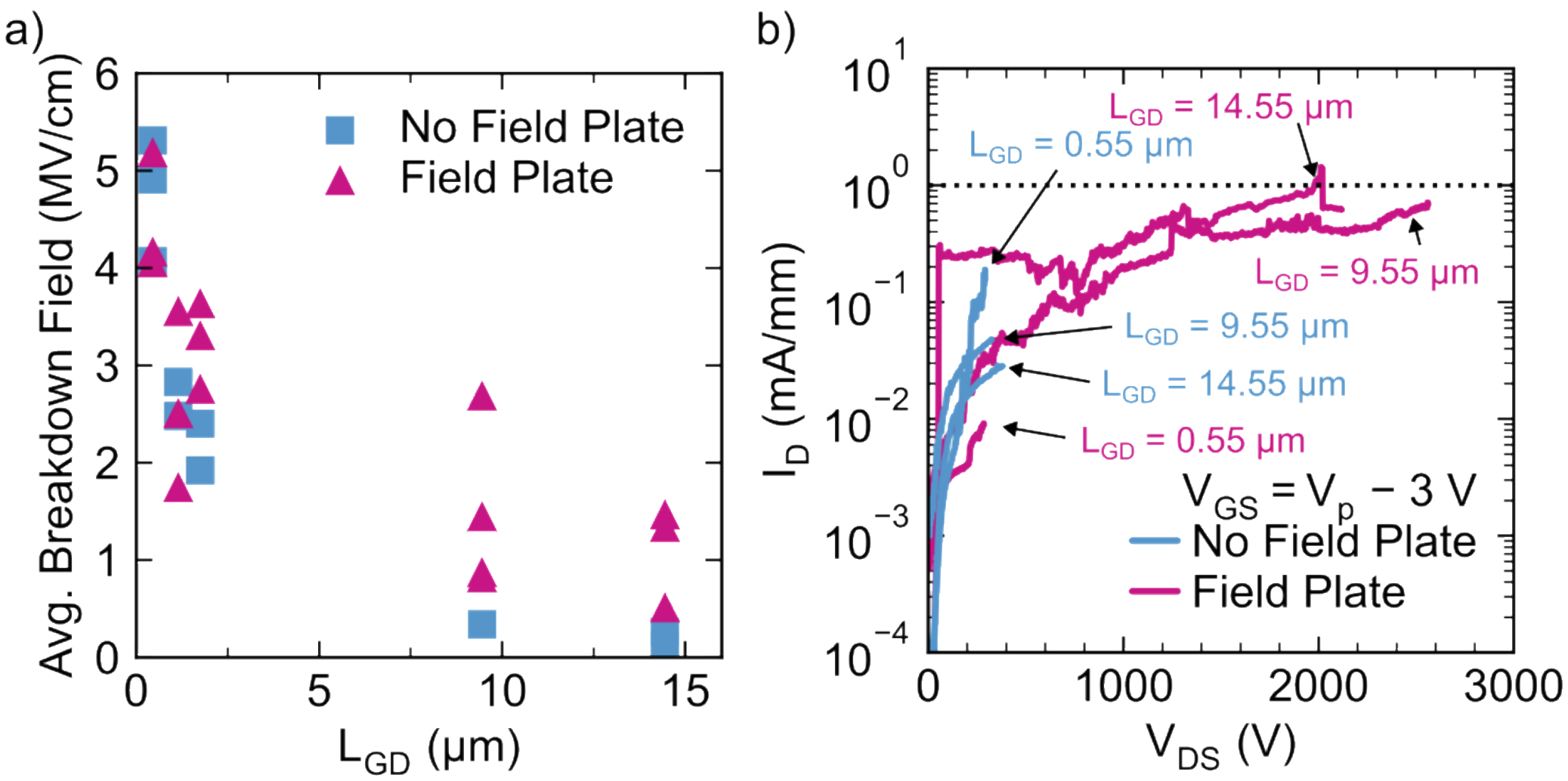

Figure 6. (a) $F_{BR}$ of devices fabricated with $Cr_2O_3$ gates and (b) three-terminal off-state breakdown measurements of devices with $Cr_2O_3$ gates with (purple) and without (blue) field plate.

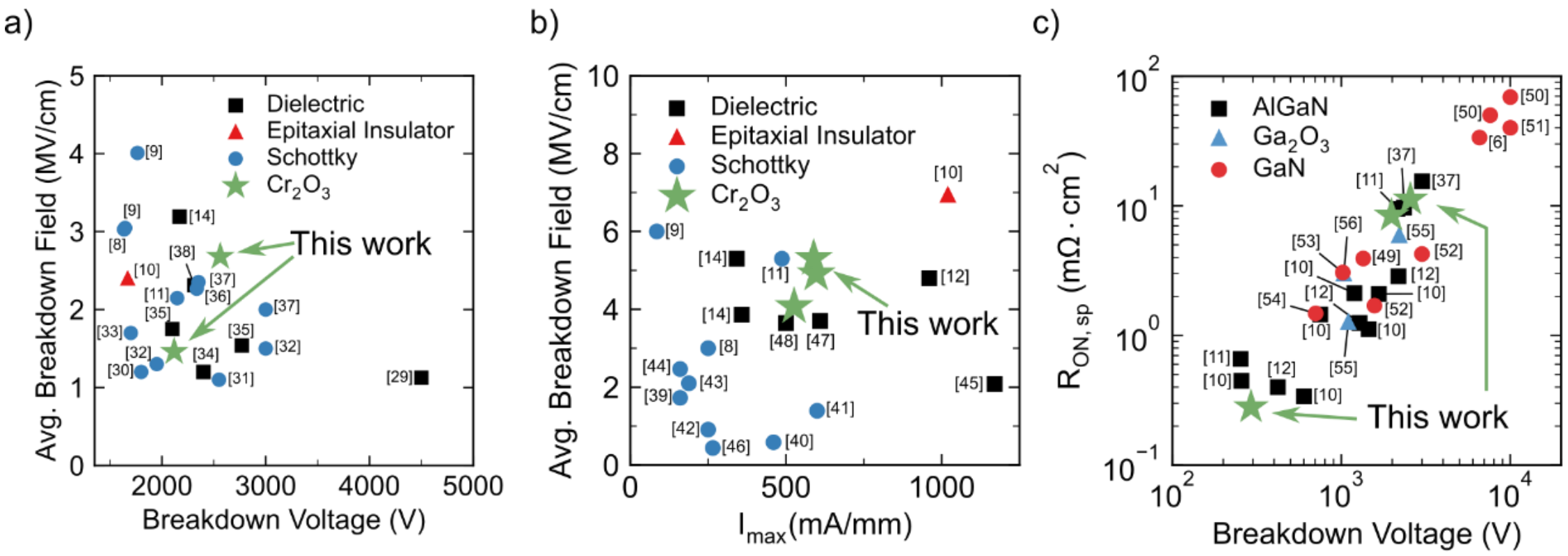

Figure 7. (a) Benchmark of $F_{BR}$ vs. $I_{max}$ for UWBG AlGaN channel transistors, (b) benchmark of $F_{BR}$ vs. $V_{BR}$ for UWBG AlGaN channel transistors, (c) benchmark of the power figure of merit $V_{BR}^2/R_{ON,sp}$ for selected state-of-the-art AlGaN, GaN, and $Ga_2O_3$ devices.

Off-state device breakdown was measured at various $L_{GD}$ up to breakdown voltage $V_{BR}$ (Fig. 6(a), Fig. 6(b)). The breakdown field is defined as the average field ($F_{BR}$) across $L_{GD}$ at the breakdown voltage ($V_{BR}$). $V_{BR}$ is defined as the voltage at which devices experienced destructive breakdown (hard breakdown) or reached drain current $I_D$ = 1 mA/mm (soft breakdown). Gate-drain spacings were defined from the edge of the conformally sputtered $Cr_2O_3$. Prior to passivation and field plate deposition, $Cr_2O_3$-gated devices measured $V_{BR}$ = 292 V, 324 V, and 381 V at $L_{GD}$ = 0.55 µm, 9.55 µm, and 14.55 µm, respectively. An average breakdown field $F_{BR}$ = 5.31 MV/cm was extracted from the $L_{GD}$ = 0.55 µm device. Low-voltage breakdown at larger $L_{GD}$ is believed to be caused by the combination of lateral depletion from the $Cr_2O_3$ sidewall and vertical field crowding in the $Cr_2O_3$ at the $Cr_2O_3$/n-AlGaN interface on the drain side. The $Cr_2O_3$ breaks down due to the large field generated from the depletion. Fig. 7(a) shows $F_{BR}$ of several high-voltage UWBG AlGaN devices[8–11,29–38]. After passivation and field plate deposition, devices with larger $L_{GD}$ showed $V_{BR}$ enhancement up to 2562 V hard breakdown at $L_{GD}$ = 9.55 µm ($V_{BR}^2/R_{ON,sp}$ = 586 MW/cm$^2$), the highest $F_{BR}$ measured at $V_{BR}$ > 2500 V, and 1980 V soft breakdown at $L_{GD}$ = 14.55 µm ($V_{BR}^2/R_{ON,sp}$ = 466 MW/cm$^2$). In Fig. 7(b), the product $F_{BR} \times I_{max}$ is shown for a selection of UWBG AlGaN devices[8–12,14,39–48]. Benchmarking $Cr_2O_3$-gated devices using the product $F_{BR} \times I_{max}$ shows superior performance to UWBG AlGaN devices with gate dielectrics, demonstrating successful breakdown enhancement using room-temperature sputtered $Cr_2O_3$. In Fig. 7(c), the selected $Cr_2O_3$-gated devices are benchmarked using the power figure of merit ($V_{BR}^2/R_{ON,sp}$) from state-of-the-art AlGaN, GaN, and $Ga_2O_3$ devices[6,10–12,37,49–56].

In conclusion, we demonstrated UWBG AlGaN PolFETs with breakdown voltage exceeding 2.5 kV at $L_{GD}$ = 9.55 μm by incorporating a room-temperature sputtered p-type $Cr_2O_3$ p-type oxide for device gates. Fabricated PolFETs with average breakdown fields higher than 5.3 MV/cm, 0.28 mΩ·cm$^2$ specific on-resistance, and 590 mA/mm maximum current density were observed ($L_{GD}$ = 0.55 μm). The P-N heterojunction gates displayed a positive gate turn-on voltage $V_{ON}$ > 3.5 V and a positive threshold voltage shift $\Delta V_{TH}$ = +1.63 V. Reverse leakage was measured below 1.5×10$^{-5}$ mA/mm ($I_{ON}/I_{OFF}$ ~ 3×10$^7$). The results reveal the potential of P-N heterojunction gate design for UWBG AlGaN channel RF and lateral power transistors while showcasing the incorporation of p-type oxides with UWBG III-nitride devices.

This work was funded by ARO DEVCOM under Grant No. W911NF2220163 (UWBG RF Center, program manager Dr. Tom Oder). This article has been co-authored by employees of National Technology & Engineering Solutions of Sandia, LLC under Contract No. DE-NA0003525 with the U.S. Department of Energy (DOE). The employee owns all right, title and interest in and to the article and is solely responsible for its contents. The United States Government retains and the publisher, by accepting the article for publication, acknowledges that the United States Government retains a non-exclusive, paid-up, irrevocable, world-wide license to publish or reproduce the published form of this article or allow others to do so, for United States Government purposes. The DOE will provide public access to these results of federally sponsored research in accordance with the DOE Public Access Plan https://www.energy.gov/downloads/doe-public-access-plan.

**Author Declarations**

**Conflict of Interest**

The authors have no conflicts to disclose.

**Data Availability**

The data that support the findings of this study are available within the article.